\documentstyle[aps,amsmath]{revtex}

\begin{document}

\draft

\title{\bf LOCALIZATION LENGTH IN DOROKHOV'S MICROSCOPIC MODEL OF
MULTICHANNEL WIRES}
\author{J.Heinrichs}
\address{Institut de physique, B5, Universit\'{e} de Li\`{e}ge, Sart
Tilman, B-4000 Li\`{e}ge, Belgium}
\maketitle
\date{\today}
\maketitle

\begin{abstract}
\thispagestyle{empty}
\noindent
We derive exact quantum expressions for the localization length $L_c$ for
weak disorder in two- and three chain tight-binding
systems coupled by random nearest-neighbour interchain hopping terms and
including random energies of the atomic sites.
These quasi-1D systems are the two- and three channel versions of
Dorokhov's model of localization in a wire of $N$
periodically arranged atomic chains.  We find that $L^{-1}_c=N.\xi^{-1}$
for the considered systems with $N=(1,2,3)$, where
$\xi$ is Thouless' quantum expression for the inverse loca\-lization length
in a single 1D Anderson chain, for weak disorder.
The inverse localization length is defined from the exponential decay of
the two-probe Landauer conductance, which is
determined from an earlier transfer matrix solution of the Schr\"{o}dinger
equation in a Bloch basis.  Our exact expressions above
differ qualitatively from Dorokhov's localization length identified as the
length scaling parameter in his scaling description
of the distribution of the participation ratio.  For $N=3$ we also discuss
the case where the coupled chains are arranged on a
strip rather than periodically on a tube.  From the transfer matrix
treatment we also obtain reflection coefficients matrices
which allow us to find mean free paths and to discuss their relation to
localization lengths in the two- and three channel
systems.

\pacs{72.15.-Rn, 73.21.+m, 73.23.Hb, 73.63.Nm - e-mail: J.Heinrichs@ulg.ac.be}
\end{abstract}

\pagestyle{plain}
\setcounter{page}{1}

\section{INTRODUCTION}

The localization length ($L_c$) of the quantum states is a fundamental
parameter in mesoscopic physics.  In particular, for
quasi-1D disordered systems (wires) of finite length $L$, it not only sets
the scale beyond which the electron states are
effectively  localized but it also determines the domain

\begin{equation}\label{eq1}
\ell\leq L\leq L_c\quad ,
\end{equation}

\noindent
in which the conductance $g_L$ displays classical Ohmic behaviour,
$g_L\propto L^{-1}$, corresponding to a diffusive metallic
regime.  Here $\ell$ denotes the elastic mean free path and

\begin{equation}\label{eq2}
L_c\simeq N\;\ell\quad ,
\end{equation}

\noindent
where $N\propto \sqrt{A}$ is the number of scattering channels in a wire of
cross-sectional area $A$.  The metallic domain
(\ref{eq1}) does not exist for 1D chains ($N=1$) and for real wires with
$N>>1$ it requires the resistance to be less than
some relatively large threshold value.  These fundamental results have
first been established by Thouless\cite{1} and are
reviewed in\cite{2,3}.\\

On the other hand, the notion of scattering channels itself is important
since it has permitted the generalization of the
well-known scaling equation for the evolution of the distribution of
resistance (conductance) as a function of length in a 1D
chain\cite{4}, in terms of the Dorokhov-Mello-Pereyra-Kumar (DMPK)
equation\cite{5,6} for a distribution of scattering
parameters related to the conductance in quasi-1D systems.  The DMPK
equation together with the numerous results derived from
it has been reviewed in\cite{7,8,9}.\\

Despite the important role played by the localization length in
multi-channel disordered systems, microscopic analytic studies
of it have remained scarce.  This is the more surprising as $L_c$ is
actually an intrinsically microscopic quantum
parameter of fundamental importance, as recalled above.  Note also that a
first principles derivation of a relation between
the localization length and the mean free path such as (\ref{eq2}) would
require separate calculation of both quantities in a
disordered atomistic multi-channel system.\\

Some years ago Dorokhov\cite{10} has discussed a solvable model of
multi-channelloca\-lization consisting of $N$ random
tight-binding chains coupled by random nearest-neighbour interchain hopping
terms and having random site-energies.  Dorokhov's
aim was to relax the assumption of isotropy of scattering parameters
underlying the derivation of the DMPK
equation\cite{5,6,7} by replacing it by the weaker assumption of equivalent
scattering channels\cite{8}.  After a fairly
sophisticated analysis, which we find difficult to follow, Dorokhov arrives
at an evolution equation for the distribution of
scattering variables (participation ratio\cite{11}) which involves a single
microscopically defined scaling parameter, which
he identifies with the localization length\cite{10}.  Dorokhov's expression
for the localization length for weak disorder is
independent of the number of channels [Equation (12) below], which seems
surprising!  On the other hand, the popular models
of transport and localization in multichannel wires such as the Thouless
tunnel-junction model \cite{1,2}, the random matrix
and maximum entropy models \cite{5,6,7,8} and the non linear sigma model
\cite{9}, do not address detailed (discrete)
microscopic models with specified disorder.\\

In a recent paper\cite{12} hereafter referred to as I, the author has
derived exact analytical expressions
for localization lengths for two- and three chain tight-binding systems
with random site energies but constant
nearest-neighbour interchain (transverse) and intrachain (longitudinal)
hopping parameters, for weak disorder.  In this model
the channels are generally non equivalent, being associated with distinct
channel-wavenumbers in the absence of
disorder\cite{12}.  The localization length is defined, as usual, by the
rate of exponential decay of the conductance.  The
conductance is determined using a transfer matrix approach for obtaining
the amplitude transmission coefficients entering into
the multi-channel Landauer formula.\\

Motivated by our doubts about the correctness of Dorokhov's result (which,
in particular, is incorrect for a 1D chain), we
have reconsidered the calculation of the localization length for the case
of two- and three equivalent channels in his model,
using the exact transfer matrix method for weak disorder developed in I.
In view of the importance of Dorokhov's miscoscopic
model in the context of scaling theories for probability distributions of
transport parameters in quasi-1D systems it seems
important to dispose of an accurate independent description of the
localization length.  On the other hand, the related
analysis of reflection matrices will allow us to calculate mean free paths
for the two- and three channel systems in the Born
approximation and thus to test Eq. (\ref{eq2}).\\

In Sect.~II we recall the Schr\"{o}dinger tight-binding equations for
Dorokhov's model for the case of two- and three chain
systems.  Dorokhov's model corresponds to periodic boundary conditions for
the chains i.e. it describes equidistant chains
arranged parallel to the axis on a tube.  We also consider an alternative
three-chain model with the parallel chains arranged
on a planar strip, which corresponds to using free boundary conditions for
the chains which are now non equivalent.  In
Sect.~III we summarize the main points of the determination of the
transfer- and scattering matrices in these models.  Using
the same parameterization of the matrix elements as in I, we limit
ourselves to the definitions of these parameters in terms of
the random site energies and hopping rates in Dorokhov's model, referring
to I for presentation of the explicit forms of the
corresponding matrices.  The results and some concluding remarks are
discussed in Sect.~IV.  In particular, we allude to a
recently studied\cite{13} weakly disordered multichain model including both
interchain and intrachain nearest-neighbour random
hopping but no site energy disorder.  This model generalizes a well-known
1D random hopping tight-binding model in which a
delocalization transition has been found at the band centre\cite{14}.  We
give an exact expression for the localization length
in this 1D model, which readily reveals the delocalization transition in
the middle of the energy band.

\section{MICROSCOPIC MULTI-CHANNEL MODEL}

The $N$-chain Dorokhov model\cite{10} of a wire consists of parallel linear
chains of $N_L$ disordered sites each (of spacing
$a=1$ and length $L=N_L a$) connected at both ends to semi-infinite ideal
(non-disordered) chains constituting the leads.  The
sites on a given chain with its associated non-disordered parts are
labelled by integers $1\leq m\leq N_L$ in the disordered
region and by $m\leq 1$ and $m\geq N_L$ in the non-disordered ones,
respectively.  The disordered chains are coupled to each
other by random hopping rates (transverse hopping) with vanishing mean
values and, correspondingly the non-disordered chains
are decoupled.  The system is described by the tight-binding
Schr\"{o}dinger equation

\begin{subequations}\label{eq3}
\renewcommand{\theequation}
{\theparentequation.\alph{equation}}
\begin{eqnarray}
\psi^i_{n+1}+\psi^i_{n-1}+\sum^N_{j=1}\varepsilon^{ij}_n\psi^j_n=E\psi^i_n,i=1,2
,\ldots\;N
&,& 1\leq n\leq N_L\quad .\\
\psi^i_{n+1}+\psi^i_{n-1}=E\psi^i_n
&,&  n < 1 \;\text{or}\; n > N_L\quad .
\end{eqnarray}
\end{subequations}

\noindent
Here $E$ is the energy and $\psi^i_m$ denotes the amplitude of the
wavefunction at a site $m$ on the $i$th chain;
$\varepsilon^{ii}_n\equiv\varepsilon^i_n$ is the random energy at a site
$n$ on chain $i$ while

\begin{equation}\label{eq4}
\varepsilon^{ij}_n=\varepsilon^{ji}_n\quad ,
\end{equation}

\noindent
is a random symmetric hopping parameter between a site $n$ on chain $i$ and
the corresponding nearest-neighbour site $n$ on
chain $j$.  The above energies, including $E$, are measured in units of a
fixed nearest-neighbour matrix element for hopping
along the individual chains (longitudinal hopping).  The random site
energies and hopping parameters are assumed to be
identically distributed independent gaussian variables with vanishing mean
and correlation
$[\varepsilon^2_0\equiv(\varepsilon_0)^2]$

\begin{subequations}\label{eq5}
\renewcommand{\theequation}
{\theparentequation.\alph{equation}}
\begin{eqnarray}
\langle\varepsilon^i_n\;\varepsilon^j_m \rangle
&=& \varepsilon^2_0\;\delta_{i,j}\delta_{m,n}\\
\langle\varepsilon^{ij}_n\;\varepsilon^{pq}_m\rangle
&=&
\varepsilon^2_0\;\delta_{m,n}(\delta_{i,p}\delta_{j,q}+\delta_{i,q}\delta_{j,p})
\quad .
\end{eqnarray}
\end{subequations}

\noindent
We note that Eqs.~(\ref{eq3}.a) describe a collection of coupled chains of
fixed separation, $a$, arranged parallel to the axis
on a tube, which corresponds to periodic boundary conditions (pbc) for the
chains.  In the absence of disorder the chains are
independent and equivalent and (\ref{eq3}.a) shows that they all couple in
the same way to the disorder.  Therefore these
independent chains define $N$ equivalent scattering channels\cite{10}.\\

We now specialize to the cases of two- and three chain systems which are
the object of this paper.  For $N=2$ and $N=3$
Eq.~(\ref{eq3}.a) may be written

\begin{equation}\label{eq6}
\begin{pmatrix}
\psi_{n+1}^1+\psi_{n-1}^1\\
\psi_{n+1}^2+\psi_{n-1}^2
\end{pmatrix}=
\begin{pmatrix}
E-\varepsilon^1_n & -\varepsilon^{12}_n\\
-\varepsilon^{21}_n & E-\varepsilon^2_n
\end{pmatrix}
\begin{pmatrix}
\psi^1_n\\ \psi^2_n
\end{pmatrix}, N=2
\quad ,
\end{equation}

\begin{equation}\label{eq7}
\begin{pmatrix}
\psi_{n+1}^1+\psi_{n-1}^1\\
\psi_{n+1}^2+\psi_{n-1}^2\\
\psi_{n+1}^3+\psi_{n-1}^3
\end{pmatrix}=
\begin{pmatrix}
E-\varepsilon^1_n & -\varepsilon^{12}_n & -\varepsilon^{13}_n\\
-\varepsilon^{21}_n & E-\varepsilon^2_n & -\varepsilon^{23}_n\\
-\varepsilon^{31}_n & -\varepsilon^{32}_n & E-\varepsilon^3_n
\end{pmatrix}
\begin{pmatrix}
\psi^1_n\\ \psi^2_n \\ \psi^3_n
\end{pmatrix}, N=3
\quad .
\end{equation}

\noindent
For completeness's sake, we also consider, for $N=3$, the case where the
parallel chains are arranged on a planar strip which
corresponds to free boundary conditions (fbc).  In this case the
Schr\"{o}dinger equation is

\begin{equation}\label{eq8}
\begin{pmatrix}
\psi_{n+1}^1+\psi_{n-1}^1\\
\psi_{n+1}^2+\psi_{n-1}^2\\
\psi_{n+1}^3+\psi_{n-1}^3
\end{pmatrix}=
\begin{pmatrix}
E-\varepsilon^1_n & -\varepsilon^{12}_n & 0\\
-\varepsilon^{21}_n & E-\varepsilon^2_n & -\varepsilon^{23}_n\\
0 & -\varepsilon^{32}_n & E-\varepsilon^3_n
\end{pmatrix}
\begin{pmatrix}
\psi^1_n\\ \psi^2_n \\ \psi^3_n
\end{pmatrix}
\quad .
\end{equation}

\noindent
Clearly, in this case, the channels are non-equivalent, but nevertheless
well-defined.\\

As in I, we shall determine the inverse localization length from the rate
of exponential decay of the conductance of the
disordered wires\cite{1,2,15},

\begin{equation}\label{eq9}
{ 1\over L_c}=-\lim_{N\rightarrow\infty}{1\over 2N}\langle \ln g\rangle
\quad ,
\end{equation}

\noindent
where averaging over the disorder may be used, as usual, because of the
self-averaging property of $\ln g$.  The conductance
is given by the Landauer two-probe conductance formula\cite{2,3},

\begin{equation}\label{eq10}
g={2e^2\over h} Tr(\hat t\hat t^+)
\quad ,
\end{equation}

\noindent
where $\hat t$ is the transmission matrix

\begin{equation}\label{eq11}
\hat t=
\begin{pmatrix}
t_{11} & t_{12} & \ldots & t_{1N}\\
t_{21} & \ldots & \ldots & \ldots\\
t_{N1} & t_{N2} & \ldots & t_{NN}
\end{pmatrix}
\quad ,
\end{equation}

\noindent
where $t_{ij}$ denotes the amplitude transmitted in channel $i$ at one end
of the wire when there is an incident amplitude in
channel $j$ at the other end.\\

We close this Section by recalling the result for the localization length
obtained by Dorokhov\cite{10} for an $N$-channel
wire described by (\ref{eq3}.a).  In the notation of (\ref{eq3}.a) and
(\ref{eq5}.a,b) it reads

\begin{equation}\label{eq12}
L_c={4-E^2\over 2\varepsilon^2_0}
\quad ,
\end{equation}

\noindent
which is independent of $N$.  This surprising result follows by combining
the expression for the localization length obtained
from the scaling equation for the distribution of the participation ratio
in the first equality of (6.26) in Ref.~\cite{10},
with the definitions (2.9), (2.8) and (2.2).  The Eq.~(\ref{eq12}) will be
discussed further in Sect.~IV.

\section{SUMMARY OF ANALYSIS}

Here we summarize the analytic study of transfer and scattering matrices
for weakly disordered two-and three channel systems
of I, as applied to the model of Sect.~II.  From this we obtain explicit
results for the various intra- and interchannel
transmission and reflection coefficients which we shall use for finding the
localization length (\ref{eq9}) and, as a check of
our results, for verifying explicitly the current conservation property.
The choice of similar notations to those used in I
will allow us, conveniently, to refer to I for the explicit forms of the
above matrices.

\subsection{Transfer matrices}

Transfer matrices, $\tilde Y_n$, for thin slices enclosing only a single
site $n$ per channel of the system described by
(\ref{eq6}-\ref{eq8}) are defined by rewriting these equations in the form

\begin{equation}\label{eq13}
\begin{pmatrix}
\psi^1_{n+1}\\
\psi^1_n\\
\psi^2_{n+1}\\
\psi^2_n\\
\vdots
\end{pmatrix}=\tilde Y_n
\begin{pmatrix}
\psi^1_n\\
\psi^1_{n-1}\\
\psi^2_n\\
\psi^2_{n-1}\\
\vdots
\end{pmatrix}
\quad ,
\end{equation}
\noindent
where

\begin{equation}\label{eq14}
\tilde Y_n\equiv\tilde X_{0n}=
\begin{pmatrix}
E-\varepsilon^1_n & -1 & -\varepsilon^{12}_n & 0\\
1 & 0 & 0 & 0\\
-\varepsilon^{21}_n & 0 & E-\varepsilon^2_n & -1\\
0 & 0 & 1 & 0
\end{pmatrix},
\varepsilon^{12}_n=\varepsilon^{21}_n
\quad ,
\end{equation}

\noindent
for $N=2$, and

\begin{equation}\label{eq15}
\tilde Y_n=
\begin{pmatrix}
E-\varepsilon^1_n & -1 & -\varepsilon^{12}_n & 0 & -\chi^{13}_n & 0\\
1 & 0 & 0 & 0 & 0 & 0\\
 -\varepsilon^{21}_n & 0 & E-\varepsilon^2_n & -1 & -\varepsilon^{23}_n & 0\\
0 & 0 & 1 & 0 & 0 & 0\\
-\chi^{31}_n & 0 & -\varepsilon^{32}_n & 0 & E-\varepsilon^3_n & -1\\
0 & 0 & 0 & 0 & 1 & 0
\end{pmatrix}
\quad ,
\end{equation}

\noindent
for $N=3$, with $\varepsilon_n^{ij}=\varepsilon_n^{ji}$, and

\begin{equation}\label{eq16}
\tilde Y_n\equiv\tilde X'_n\;,\;\chi^{ij}_n=0\;\text{for fbc}
\quad ,
\end{equation}

\begin{equation}\label{eq17}
\tilde Y_n\equiv\tilde
X^{\prime\prime}_n\;,\;\chi^{ij}_n=\varepsilon^{ij}_n\;\text{for pbc}
\quad .
\end{equation}

In order to study the reflection and transmission of plane waves by a
disordered wire we must use a basis corresponding to
waves propagating independently from left to right and from right to left
in the absence of disorder.  Such a basis is
provided by the Bloch waves supported by the system (\ref{eq3}.a,b) for
vanishing disorder.  The Bloch waves are the solutions
for real $k$ of the eigenvalue equation

\begin{equation}\label{eq18}
\tilde Y_0
\begin{pmatrix}
\psi^1_{n,\pm}\\
\psi^1_{n-1,\pm}\\
\psi^2_{n,\pm}\\
\psi^2_{n-1,\pm}\\
\vdots
\end{pmatrix}=
e^{\pm ik}
\begin{pmatrix}
\psi^1_{n,\pm}\\
\psi^1_{n-1,\pm}\\
\psi^2_{n,\pm}\\
\psi^2_{n-1,\pm}\\
\vdots
\end{pmatrix}
\quad ,
\end{equation}

\noindent
where $\tilde Y_0\equiv\tilde X_{00},\tilde X'_0,\tilde X^{\prime\prime}_0$
denotes the transfer matrices (\ref{eq14}) and
(\ref{eq15}-\ref{eq17}) in the absence of disorder i.e.
$\varepsilon^1_n=\varepsilon^2_n=\varepsilon^3_n=\varepsilon^{ij}_n=0$.
The wavenumbers $k$ are given by

\begin{equation}\label{eq19}
2\cos k=E
\quad ,
\end{equation}

\noindent
for energies restricted to the band $-2\leq E\leq 2$.  For definiteness we
choose $0\leq k\leq\pi$ so that the eigenfunctions

\begin{equation}\label{eq20}
\psi^j_{n,\pm}\sim e^{\pm ink}
\quad ,
\end{equation}

\noindent
correspond to Bloch waves travelling from left to right and from right to
left, respectively.\\

In the absence of disorder, the transfer matrices (\ref{eq14}) and
(\ref{eq15}) are diagonalized in the basis of Bloch wave
amplitudes (\ref{eq20}).  In transforming (\ref{eq13}) to the Bloch wave
basis and performing, in particular, the
corresponding similarity transformation (defined by the matrix $\widehat W$
of the eigenvectors in (\ref{eq18})\cite{12}) of
the transfer matrix $\tilde Y_n$ for $N=2$ and $N=3$, respectively, we use
the same parameterization for $\widehat
Y_n\equiv\widehat W^{-1}\tilde Y_n\widehat W$ as in Eqs (\ref{eq22})
($N=2$) and (\ref{eq23}) ($N=3$) of I (where we now put
$k_1=k_2=k_3\equiv k$).  For the models (\ref{eq6}-\ref{eq8}) the
parameters introduced in I are found to be given by:

\begin{equation}\label{eq21}
a_{1n}={\varepsilon^1_n\over 2\sin k}\;,\; a_{2n}={\varepsilon^2_n\over
2\sin k}\;,\; b_n=-{\varepsilon^{12}_n\over 2\sin k}
\quad ,
\end{equation}

\noindent
for $N=2$,

\begin{multline}\label{eq22}
a_{1n}={\varepsilon^1_n\over 2\sin k}\;,\; a_{3n}={\varepsilon^3_n\over
2\sin k}\;,\; b_{2n}={\varepsilon^2_n\over 2\sin k}
\;,\\
c_n=f_n={\varepsilon^{12}_n\over 2\sin k}\;,\;
d_n=q_n={\varepsilon^{23}_n\over 2\sin k}\;,\; g_n=p_n=0
\quad ,
\end{multline}

\noindent
for $N=3$ with fbc,

\begin{multline}\label{eq23}
a_{1n}={\varepsilon^1_n\over 2\sin k}\;,\; a_{3n}={\varepsilon^3_n\over
2\sin k}\;,\; b_{2n}={\varepsilon^2_n\over 2\sin k}
\;,\\
c_n=f_n={\varepsilon^{12}_n\over 2\sin k}\;,\;
g_n=p_n={\varepsilon^{13}_n\over 2\sin k}\;,\;
d_n=q_n={\varepsilon^{23}_n\over 2\sin k}
\quad ,
\end{multline}

for $N=3$ with pbc.

Finally, the transfer matrices of the disordered wires of length $L=N_La$
are the products of Bloch wave transfer matrices
associated with the $N_L$ individual thin slices $n$,

\begin{equation}\label{eq24}
\widehat Y_L=\prod^{N_L}_{n=1}\widehat Y_n
\quad .
\end{equation}

\noindent
For weak disorder it is sufficient to explicitate (\ref{eq24}) to linear
order in the random energies $\varepsilon^i_n$ and
$\varepsilon_n^{ij}$ for the purpose of studying averages to lowest order
in the correlations (\ref{eq5}.a,b).  These
correlations imply indeed that different slices in (\ref{eq24}) are
uncorrelated.  The Bloch wave transfer matrices are given
explicitly by Eqs (\ref{eq30}) ($N=2$) and (\ref{eq32}) ($N=3$) of I, with
the parameters defined in (\ref{eq21}-\ref{eq23})
above and the wavenumbers $k_1,k_2,k_3$ replaced by $k$ in (\ref{eq19}).

\subsection{Scattering matrices}

The scattering of plane waves (reflection and transmission) at and between
the two ends of the random quasi-1D systems is governed by the $S$-matrix,

\begin{equation}\label{eq25}
\widehat S=
\begin{pmatrix}
\hat r^{-+} & \hat t^{--}\\
\hat t^{++} & \hat r^{+-}
\end{pmatrix}\quad ,
\end{equation}

\noindent
where

\begin{equation}\label{eq26}
\hat t^{\mp\mp}=
\begin{pmatrix}
 t^{\mp\mp}_{11} &  t^{\mp\mp}_{12} & \cdots\\
 t^{\mp\mp}_{21} &  t^{\mp\mp}_{22} & \cdots\\
\vdots & \vdots & \vdots
\end{pmatrix}\quad ,
\end{equation}

\noindent
and

\begin{equation}\label{eq27}
\hat r^{\pm\mp}=
\begin{pmatrix}
 r^{\pm\mp}_{11} &  r^{\pm\mp}_{12} & \cdots\\
 r^{\pm\mp}_{21} &  r^{\pm\mp}_{22} & \cdots\\
\vdots & \vdots & \vdots
\end{pmatrix}\quad .
\end{equation}

\noindent
Here $t_{ij}^{++} (t_{ij}^{--})$ and $r_{ij}^{-+}(r_{ij}^{+-})$ denote the
transmitted and reflected amplitudes in channel $i$ when there is a unit
flux incident from the left (right) in channel $j$.  Left to right- and
right to left directions are labelled + and -, respectively.  The
$S$-matrix expresses outgoing wave amplitudes in terms of ingoing ones on
either side of the quasi-1D disordered wire via the scattering
relations

\begin{equation}\label{eq28}
\begin{pmatrix}
0 \\ 0'
\end{pmatrix}=
\widehat S
\begin{pmatrix}
I \\ I'
\end{pmatrix}\quad .
\end{equation}

\noindent
Here $I$ and $I'$ (0 and $0'$) denote ingoing (outgoing) amplitudes at the
left and right sides of the disordered region, respectively.  It follows
from current conservation that e.g. for a unit flux which is incident from
the right in channel $i$ one has

\begin{equation}\label{eq29}
\sum^{N}_{j=1}(\mid t^{--}_{ji}\mid^2+\mid r^{-+}_{ji}\mid^2)=1\quad .
\end{equation}

\noindent
Likewise, one has also

\begin{equation*}
\sum^{N}_{j=1}(\mid t^{++}_{ji}\mid^2+\mid r^{+-}_{ji}\mid^2)=1\quad.
\hspace{6cm}\text{(29.a)}
\end{equation*}

As shown in I, the wavefunction amplitudes at sites $n$ and $n-1$ in the
Bloch representation correspond to waves travelling
from left to right and from right to left, respectively.  Like in I, we
thus rename the wavefunction amplitudes in the Bloch
representation of the transfer equations (\ref{eq13}),

\begin{equation}\label{eq30}
\widehat W^{-1}
\begin{pmatrix}
\psi^1_{n+1}\\
\psi^1_n\\
\psi^2_{n+1}\\
\psi^2_n\\
\vdots
\end{pmatrix}=
\widehat Y_n\widehat W^{-1}
\begin{pmatrix}
\psi^1_n\\
\psi^1_{n-1}\\
\psi^2_n\\
\psi^2_{n-1}\\
\vdots
\end{pmatrix}
\quad ,
\end{equation}

\noindent
by defining e.g.

\begin{equation}\label{eq31}
\widehat W^{-1}
\begin{pmatrix}
\psi^1_n\\
\psi^1_{n-1}\\
\psi^2_n\\
\psi^2_{n-1}\\
\vdots
\end{pmatrix}
\equiv
\begin{pmatrix}
a^+_{1,n-1}\\
a^-_{1,n-1}\\
a^+_{2,n-1}\\
a^-_{2,n-1}\\
\vdots
\end{pmatrix}
\quad .
\end{equation}

\noindent
Likewise, using a similar notation for wave amplitudes transferred from
$n=0$ to $n=N_L$ across the disordered wire of length
$L=N_La$, we write the corresponding wave transfer equation in the Bloch
representation, which follows by iterating
(\ref{eq30}), in the form

\begin{equation}\label{eq32}
\begin{pmatrix}
a^+_{1,L}\\
a^-_{1,L}\\
a^+_{2,L}\\
a^-_{2,L}\\
\vdots
\end{pmatrix}
=\widehat Y_L
\begin{pmatrix}
a^+_{1,0}\\
a^-_{1,0}\\
a^+_{2,0}\\
a^-_{2,0}\\
\vdots
\end{pmatrix}
\quad ,
\end{equation}

\noindent
The components of the out- and ingoing waves column vectors in (\ref{eq28})
are thus
$a^-_{1,0},a^-_{2,0},\ldots a^-_{N,0},a^+_{1,L},a^+_{2,L},\ldots a^+_{N,L}$
and
$a^+_{1,0},a^+_{2,0},\ldots a^+_{N,0},a^-_{1,L},a^-_{2,L}\ldots a^-_{N,L}$,
respectively.  With the so defined vectors of
outgoing and incoming amplitudes, the $S$-matrix is obtained by rearranging
the equation (\ref{eq32}) so as to bring them in
the form (\ref{eq28}).  The details of this somewhat lengthy calculation
are explicitated in I.  The explicit forms of the
scattering matrices, Eqs (46-47) and (48,48.a-48.f) of I, for $N=2$ and $N=3$,
respectively, are expressed in terms of transfer matrix elements which are
themselves defined in terms of general parameters
given by (\ref{eq21}-\ref{eq23}) above in the case of Dorokhov's model.
These $S$-matrices readily yield the transmission and
reflection submatrices in (\ref{eq25}).

\section{RESULTS AND CONCLUDING REMARKS}

The explicit  expressions of the transmission- and reflection coefficients,
$|t^{--}_{ij}|^2$ and $|r^{-+}_{ij}|^2$, in terms
of the general parameters defining the transfer matrices in I are given in
an appendix in I.  By inserting the present
parameter values (\ref{eq22}) and (\ref{eq23}) for the two- and
three-channel Dorokhov models in these expressions and
averaging over the disorder, using (\ref{eq5}.a) and (\ref{eq5}.b), we
obtain the following results, exact to order
$\varepsilon^2_0$:

\begin{equation}\label{eq33}
\langle |t^{--}_{11}|^2 \rangle=\langle |t^{--}_{22}|^2 \rangle=1-{3
N_L\varepsilon^2_0\over4\sin^2 k}
\quad ,
\end{equation}

\begin{equation}\label{eq34}
\langle |t^{--}_{12}|^2 \rangle=\langle |t^{--}_{21}|^2
\rangle={N_L\varepsilon^2_0\over4\sin^2 k}
\quad ,
\end{equation}

\begin{equation}\label{eq35}
\langle |r^{-+}_{ij}|^2 \rangle={N_L\varepsilon^2_0\over4\sin^2 k}\;,\;
i,j=(1,2)
\quad ,
\end{equation}

\noindent
for $N=2$,

\begin{equation}\label{eq36}
\langle |t^{--}_{11}|^2 \rangle=\langle |t^{--}_{33}|^2 \rangle=1-{3
N_L\varepsilon^2_0\over4\sin^2 k}
\quad ,
\end{equation}

\begin{equation}\label{eq37}
\langle |t^{--}_{22}|^2 \rangle=1-{5 N_L\varepsilon^2_0\over4\sin^2 k}\quad ,
\end{equation}

\begin{equation}\label{eq38}
\langle |t^{--}_{12}|^2 \rangle=\langle |t^{--}_{21}|^2
\rangle={N_L\varepsilon^2_0\over4\sin^2 k}
\quad ,
\end{equation}

\begin{equation}\label{eq39}
\langle |t^{--}_{23}|^2 \rangle=\langle |t^{--}_{32}|^2
\rangle={N_L\varepsilon^2_0\over4\sin^2 k}
\quad ,
\end{equation}

\begin{equation}\label{eq40}
\langle |t^{--}_{13}|^2 \rangle=\langle |t^{--}_{31}|^2 \rangle=0
\quad ,
\end{equation}

\begin{multline}\label{eq41}
\langle |r^{-+}_{11}|^2 \rangle=\langle |r^{-+}_{22}|^2 \rangle=\langle
|r^{-+}_{33}|^2 \rangle
=\langle |r^{-+}_{12}|^2 \rangle=\langle |r^{-+}_{21}|^2 \rangle=\\
\langle |r^{-+}_{23}|^2 \rangle=\langle |r^{-+}_{32}|^2
\rangle={N_L\varepsilon^2_0\over4\sin^2 k}
\quad ,
\end{multline}

\begin{equation}\label{eq42}
\langle |r^{-+}_{13}|^2 \rangle=\langle |r^{-+}_{31}|^2 \rangle=0
\quad ,
\end{equation}

\noindent
for $N=3$ with free b.c.,

\begin{equation}\label{eq43}
\langle |t^{--}_{11}|^2 \rangle=\langle |t^{--}_{22}|^2 \rangle=\langle
|t^{--}_{33}|^2 \rangle
=1-{5N_L\varepsilon^2_0\over4\sin^2 k}
\quad ,
\end{equation}

\begin{equation}\label{eq44}
\langle |t^{--}_{ij}|^2 \rangle={N_L\varepsilon^2_0\over4\sin^2 k}\;,\;
i\neq j\;,\;i,j=(1,2,3)
\quad ,
\end{equation}

\begin{equation}\label{eq45}
\langle |r^{-+}_{ij}|^2 \rangle={N_L\varepsilon^2_0\over4\sin^2
k}\;,\;i,j=(1,2,3)
\quad ,
\end{equation}

for $N=3$ with periodic b.c.

One readily verifies, as a check of the explicit results
(\ref{eq33}-\ref{eq45}), that in all cases ($N=2$, $N=3$ with fbc and
$N=3$ with pbc) the current conservation property (\ref{eq29}) is obeyed.

Next, by evaluating the averaged traces $\langle Tr[\hat t^{--}(\hat
t^{--})^+]\rangle$, successively for the three mo\-dels
using (\ref{eq33}-\ref{eq34}), (\ref{eq36}-\ref{eq40}) and
(\ref{eq43}-\ref{eq44}), respectively, we get:

\begin{eqnarray}\label{eq46}
\langle Tr[\hat t^{--}(\hat t^{--})^+] \rangle
&=& 2-{N_L\varepsilon^2_0\over \sin^2 k}\;,\; N=2\quad ,\\
&=& 3-{7N_L\varepsilon^2_0\over 4\sin^2 k}\;,\; N=3\;\text{with fbc}\quad ,\\
&=& 3-{9N_L\varepsilon^2_0\over 4\sin^2 k}\;,\; N=3\;\text{with pbc}
\quad .
\end{eqnarray}

\noindent
For the inverse localization lengths defined in (\ref{eq9}-\ref{eq10}) we
then obtain:

\begin{eqnarray}\label{eq49}
{1\over L_c}={\varepsilon^2_0\over 4\sin^2 k}\quad
&,& \quad \text{for}\; N=2\quad ,\\
{1\over L_c}={7\varepsilon^2_0\over 24\sin^2 k} \quad
&,& \quad  \text{for}\; N=3\;\text{with fbc}\quad ,\\
{1\over L_c}={3\varepsilon^2_0\over 8\sin^2 k} \quad
&,& \quad \text{for}\; N=3\;\text{with pbc}
\quad .
\end{eqnarray}

\noindent
These expressions are exact to order $\varepsilon^2_0$ for weak disorder.

\noindent
It is instructive to compare (49-51) with the localization length, $\xi$,
for weak disorder in a
one-dimensional chain with random site energies.  In this case
Thouless\cite{16} obtained the exact expression

\begin{equation}\label{eq52}
{1\over\xi}={\varepsilon^2_0\over 8\sin^2 k}
\quad ,
\end{equation}

\noindent
which has been rederived in I (see also\cite{17}) using transfer matrices.
We observe that the inverse localization lengths
for pbc in (\ref{eq49}) and (51) take the values $2/\xi$ and $3/\xi$ for
$N=2$ and for $N=3$, respectively.  The
constant value (\ref{eq12}) obtained by Dorokhov\cite{10} for arbitrary $N$
differs qualitatively from these exact results.
Using (\ref{eq19}) and (\ref{eq52}) Dorokhov's expression may be written

\begin{equation}\label{eq53}
{1\over L_c}={4\over\xi}\;,\; N\;\text{arbitrary}
\quad .
\end{equation}

In fact, the exact expressions (\ref{eq49}) and (51), together with the 1D
expression (\ref{eq52}), suggest that the
actual form for the inverse localization length for weak disorder for
arbitrary $N$ could be

\begin{equation}\label{eq54}
{1\over L_c}={N\over\xi}
\quad .
\end{equation}

\noindent
We also note, incidently, that in our analysis of the two- and
three-channel Dorokhov models the localization length reduces
precisely to the 1D result (\ref{eq52}) in the limit of no interchain
hopping ($\varepsilon^{ij}_n=0$), as expected.  Indeed,
for $\varepsilon^{ij}_n=0$, we have $t^{--}_{ij}=r^{-+}_{ij}=0$ for $i\neq
j$ and from the explicit expressions of the random
transmission coefficients in the appendix of I we get, using
(\ref{eq21}-\ref{eq23}) and (\ref{eq5}.a),

\begin{eqnarray*}
\langle Tr[\hat t^{--}(\hat t^{--})^+] \rangle
&=& 2-{N_L\varepsilon^2_0\over 2\sin^2 k}\;,\; N=2\quad ,\\
&=& 3\left(1-{N_L\varepsilon^2_0\over 4\sin^2 k}\right)\;,\; N=3
\quad ,
\end{eqnarray*}

\noindent
which both lead to $1/L_c=1/\xi$.

On the other hand, the above results for reflection coefficients may be
used for obtaining explicit expressions for mean free
paths in the few-channel quasi-1D systems.  The mean free path for an
$N$-channel wire is defined by\cite{8,18}

\begin{equation}\label{eq57}
{1\over\ell_N}={1\over N_LN}\sum_{i,j}\langle |r^{-+}_{ij}|^2\rangle
\quad .
\end{equation}

\noindent
We then obtain successively from Eqs. (\ref{eq35}), (\ref{eq41}-\ref{eq42})
and (\ref{eq45})

\begin{eqnarray}\label{eq58}
{1\over\ell_2}
&=& {\varepsilon^2_0\over 2\sin^2 k}\quad ,\\
{1\over\ell_3}
&=& {7\over 12} {\varepsilon^2_0\over \sin^2 k}\quad\text{(fbc)}\quad ,\\
{1\over\ell_3}
&=& {3\over 4} {\varepsilon^2_0\over \sin^2 k}\quad\text{(pbc)}\quad .
\end{eqnarray}

\noindent
In the one-dimensional case one gets similarly, by returning from
(\ref{eq52}) to the determination of the reflection
coefficient,

\begin{equation}\label{eq59}
{1\over\ell_1}={\varepsilon^2_0\over 4\sin^2 k}
\quad .
\end{equation}

\noindent
The expressions (56-59) correspond to the Born approximation of impurity
scattering.  By comparing (56-59) successively with
the localization lengths in (\ref{eq49}-\ref{eq52}) we find that in all cases

\begin{equation}\label{eq60}
L_c=2\ell_N\quad ,\quad N=1,2,3
\quad .
\end{equation}

\noindent
We also note that a similar calculation of mean free paths for the two- and
three channel wire models with constant
interchain hopping rates discussed in I\cite{12} also leads to Eq.
(\ref{eq60}) for $N=2$ and $N=3$.  The localization lengths
for the multichannel systems in I are given by Eqs (58), (73) and (86) of
that reference, respectively, and
the corresponding reflection coefficients entering in (\ref{eq57}) above,
are given by Eqs (52-57), (67-72) and
(83-85) of\cite{12}.

The Eq.(\ref{eq60}) for the one-dimensional case coincides with the
relation between the localization length and the mean free
path derived by Thouless from kinetic theory\cite{19}.  Our treatment thus
establishes a similar exact relationship for two-
and three-channel systems both for Dorokhov's model and for the model with
constant interchain hopping in I. The exact
universal expression (\ref{eq60}) differs qualitatively from Eq.
(\ref{eq2}) discussed earlier, mainly for $N>>1$
\cite{1,2,6,7,9,10}, and does not suggest the existence of a well-defined
diffusive (metallic) regime,
\linebreak
$\ell_N<<L<<L_c$, in few-channel systems.  We recall that in the above
references the mean free path is introduced as a fixed
length scale beyond which metallic diffusion takes place (when it is not
inhibited by localization). Our microscopic
analysis yields explicit expressions both for localization lengths and for
mean free paths.

Finally, from (50) and (51) it follows that the difference in transverse
boundary conditions for the
corresponding three-channel models has only a minor influence on the
localization lengths.

The transfer matrix approach discussed in I may also be applied for
studying the delocalization transition which has recently
been found at the band centre in weakly disor\-dered multi-chain systems
including both nearest-neighbour inter- and intrachain
random hopping terms but no site energy-disorder\cite{13}.  This
delocalization transition exists already in a one-dimensional
chain with random hopping, as has been known for some time\cite{14}.  In
this case it may be readily revealed by studying the
localization length
$L^{-1}_c=-\lim_{L\rightarrow\infty}(2L)^{-1}\langle\ln|t^{\pm\pm}|^2\rangle$ of
the chain.  Consider the
Schr\"{o}dinger equation

\begin{equation}\label{eq55}
(1+\eta_n)(\psi_{n+1}+\psi_{n-1})=E\psi_n
\quad ,
\end{equation}

\noindent
where $\eta_n$ is a gaussian random nearest-neighbour hopping parameter (with
$\langle\eta_m\eta_n\rangle=\eta^2_0\delta_{m,n}$) measured in units of the
non random hopping parameter.  From a transfer
matrix analysis of (\ref{eq55}) similar to that used for obtaining the
transmission coefficient and the corresponding
localization length (\ref{eq52}) for a weakly disordered Anderson
chain\cite{12,17} we get

\begin{equation}\label{eq56}
{1\over L_c}={\eta^2_0\over 2}{\cos^2 k\over\sin^2 k}\;,\;E=2\cos k
\quad .
\end{equation}

\noindent
This expression, which is exact to order $\eta_0^2$, displays the
divergence of the localization length in the middle of the
energy band, $E=0$.

\end{document}